# Thermodynamic driving force of transient negative capacitance of ferroelectric capacitors

*Yuanyuan Zhang, Xiaoqing Sun, Junshuai Chai, Hao Xu, Xueli Ma, Jinjuan Xiang, Kai Han, Xiaolei Wang\*, and Wenwu Wang*

Dr. Yuanyuan Zhang, Xiaoqing Sun, Junshuai Chai, Hao Xu, Prof. Xueli Ma, Jinjuan Xiang, Xiaolei Wang, Wenwu Wang
Key Laboratory of Microelectronics & Integrated Technology, Institute of Microelectronics, Chinese Academy of Sciences, and College of Microelectronics, University of Chinese Academy of Sciences, Beijing100049, China
E-mail: wangxiaolei@ime.ac.cn

Prof. Kai Han
Department of Physics and Electronic Science, Weifang University, Weifang 261061, China

This paper investigates the thermodynamic driving force of transient negative capacitance (NC) in the series circuit of the resistor and ferroelectric capacitor (*R*-FEC). We find that the widely used Landau-Khalatnikov (L-K) theory, that is, the minimum of the Gibbs free energy, is inapplicable to explain the transient NC. The thermodynamic driving force of the transient NC phenomenon is the minimum of the difference between the elastic Gibbs free energy and the electric polarization work. The appearance of the transient NC phenomenon is not due to the widely accepted view that the ferroelectric polarization goes through the negative curvature region of elastic Gibbs free energy landscape ($G_a$). Instead, the transient NC phenomenon appears when the energy barrier of $G_a$ disappears. The transient NC is dependent on both the intrinsic ferroelectric material parameters and extrinsic factors in the *R*-FEC circuit.

# 1. Introduction

Complementary metal-oxide-semiconductor field-effect transistor (MOSFET) technology[1] is the cornerstone of the contemporary information industry. Its ubiquity has been realized by the dramatic increase of its processing capacity every unit chip area[2], due to the exponential scaling of the MOSFET devices, well known as Moore's law. However, the supply voltage has not been proportionally reduced, and limited to ~0.8 V in modern integrated circuits, in order to effectively suppress the power consumption. The power consumption has been the bottleneck of scaling down the contemporary MOSFET devices[3]. The most effective way to reduce the power consumption of an integrated circuit is to reduce the supply voltage[4-6]. However, when the supply voltage is lowered, the drive current is sacrificed, which would decrease the operating speed of the transistor. The reason is that at least ~60 mV of gate voltage is needed to change the drain to source current by an order of magnitude, which is known as Boltzmann limit. In order to avoid this problem, it is necessary to have the highest possible on-off current ratio $I_{ON}/I_{OFF}$. This can be achieved by reducing the subthreshold swing (SS). For a classical MOSFET, the minimum SS is 60mV/decade at room temperature[7], because of the Boltzmann distribution of channel carriers limit. To overcome this fundamental limit, the negative capacitance (NC) FETs[8-21] have been proposed to amplify the semiconductor surface potential by the pioneering work of Salahuddin and Datta in 2008[8]. Moreover, the NC FETs are the most promising among the future steep subthreshold devices[22, 23] since the discovery of ferroelectricity in doped hafnium oxide[24-30].

So far there are two kinds of NC experimentally reported: (1) stabilized (steady-state) NC[31, 32], and (2) transient NC[33, 34]. The stabilized NC has been experimentally demonstrated in FE-DE (ferroelectric-dielectric) heterostructures and superlattices of PZT/STO, BTO/STO, SRO/STO, SRO/GSO, and SRO/DSO by total capacitance enhancement[35-37]. Moreover, by forming polarization arrays of clockwise and anticlockwise vortex-like structures, the stabilized NC was also observed in epitaxial superlattices composed of ferroelectric $PbTiO_3$ and non-ferroelectric $SrTiO_3$[38]. On the other hand, several models of steady-state NC have been proposed, associated with stabilizing the negative curvature region of ferroelectric double-well Gibbs energy landscape by adding the parabolic energy landscape of normal dielectric to the double-well landscape of ferroelectric[39]. In addition, domain wall motion has been suggested to produce NC in a multidomain system[40-43]. Nevertheless, steady-state NC suffers from big disputes and doubts[44-46]. For the case of transient NC, it has been experimentally observed in a resistor-ferroelectric capacitor (*R*-FEC) network by applying a voltage pulse[33, 42, 47-49]. The transient NC appears during the polarization switching of the ferroelectric material. Its physical origin is widely accepted as the mismatch between the free charges on the metal electrode and ferroelectric polarization[46, 48, 50, 51]. So far, the understanding of the basic physics for transient NC is based on the Landau–Ginzburg–Devonshire (L-G-D) theory and the Landau-Khalatnikov (L-K) equation[52-55]. The polarization state in this theory is described by double-well Gibbs free energy landscape as a function of the polarization *P*, where there exist two stable polarization states with the minimum energy. When a voltage pulse is applied to the ferroelectric capacitor, the polarization state will switch along the direction of Gibbs free energy minimization[55]. It is worth noting that this process is only valid under the electric field $E$=constant across the ferroelectric capacitor. However, $E$ is not a constant field but changes with time in transient NC[46, 56, 57], which will be shown in section 2. This means that the conventional L-K theory is improper to explain the transient NC phenomenon. Thus, it is necessary to reconsider the physical origin and find the thermodynamic driving force for the transient NC phenomenon. Moreover, the modern explanation of transient NC lacks a detailed physical picture, and a thorough understanding of the underlying physics is still absent. The above issues need urgent understanding because it limits NC progress and is critical for guiding the design of device structure such as ferroelectric parameters.

In this work, we revisit the thermodynamic driving force of the transient NC phenomenon in the $R$-FEC circuit. We find that the thermodynamic driving force is the minimization of a new thermodynamic function $\eta$ instead of Gibbs free energy $G$, where $\eta$ is equal to the difference between the elastic Gibbs free energy and the electric polarization work. This means that the polarization switches along the direction of $\eta$ minimization. Moreover, this viewpoint is further supported by the analysis based on experimental $P$-$E$ curves. In addition, we also find that the appearance of the transient NC phenomenon is not due to the widely accepted viewpoint that the ferroelectric polarization goes through the negative curvature region of elastic Gibbs free energy landscape ($G_a$). Instead, the NC phenomenon appears when the energy barrier of $G_a$ disappears. Moreover, the transient NC is dependent on both the intrinsic ferroelectric material parameters and extrinsic factors in the $R$-FEC circuit.

**2. Limits of Landau-Khalatnikov (L-K) equation in explaining the transient NC**

Generally, transient NC is explained by the L-K equation. The L-K equation, describes a thermodynamics process, where the FE polarization obeys the Gibbs free energy minimization, can be expressed as

$$\rho \frac{dP}{dt} = -\frac{\partial G}{\partial P} \tag{1}$$

where $\rho$ is the viscosity coefficient which means the delay of polarization $P$ with respect to the electric field. $G$ is Gibbs free energy, which can be expanded in terms of $P$ based on L-G-D theory, expressed as[58]

$$G = \alpha P^2 + \beta P^4 + \gamma P^6 - PE = G_a - PE \tag{2}$$

where $\alpha$, $\beta$, and $\gamma$ are the thermodynamic expansion coefficients, $E$ is the electric field across the FE capacitor, and $G_a$ is elastic Gibbs free energy. In L-K equation, the term $dG/dP$ can be regarded as driving force $F$ which causes polarization switching toward the local minimum of the free energy. The traditional expression of $F$ is

$$F = -\frac{\partial G}{\partial P} = E - \left(2\alpha P + 4\beta P^3 + 6\gamma P^5\right) = E - \frac{dG_a}{dP} \tag{3}$$

The above theory is widely used to explain the transient NC[42, 47], and successfully in line with the experimental results[42, 47]. It should be noted that the key of the L-K equation in explaining the transient NC is the minimization of the Gibbs free energy $G$. However, there is a precondition of using the $G$ minimization, that is, isoelectric field, which has been ignored during the previous application of the L-K equation in understanding the transient NC. Here we show why the isoelectric field is the precondition. According to the first and second laws of thermodynamics, we have $TdS \geq dU-dW$, where $T$, $S$, and $U$ are the temperature, the entropy, and the total internal energy of the system, respectively, and $W$ is the external work. In FE materials, the total differential of external work can be expressed as $dW=Xdx+EdD$, where $Xdx$ and $EdD$ are mechanic work and electric work, respectively. $x$ is the elastic strain, $X$ is the stress and $D$ is the electric displacement field. Thus, the total differential of internal energy can be given as $dU \leq TdS+Xdx+EdP+d(1/2\varepsilon_0 E^2)$. Changing the order of the last equation, it can be rewritten as

$$d(U - \frac{1}{2}\varepsilon_0 E^2) - TdS - Xdx - EdP \leq 0. \tag{4}$$

In the process of isothermal ($dT=0$) and without any stress ($dX=0$), Equation (4) is further written as

$$d(U - \frac{1}{2}\varepsilon_0 E^2) - TdS - SdT - Xdx - xdX - EdP \leq 0$$

or

$$d(U - TS - Xx - \frac{1}{2}\varepsilon_0 E^2) - EdP \leq 0. \tag{5}$$

We consider the case of assuming $E$ = constant. Correspondingly, $PdE=0$. Adding this item to the left side of the Equation (5) obtains

$$d(U - TS - Xx - EP - \frac{1}{2}\varepsilon_0 E^2) \leq 0. \tag{6}$$

Here, the items ($U$-$TS$-$Xx$-$EP$-$1/2\varepsilon_0 E^2$) in brackets is Gibbs free energy $G$. Then, Equation (6) becomes

$$dG \leq 0. \tag{7}$$

This indicates that the Gibbs free energy no longer increases at constant electric fields. Thus, the polarization switching along the direction of Gibbs free energy minimization is only applicable when keeping the electric field across FE constant. The next question is whether the transient NC process meets the isoelectric field condition. From the reported experimental *P-E* curves of transient NC measured on positive rectangular voltage pulse in the *R*-FEC circuit[42, 47], as shown in Figures 2(a) and 2(b), we can find that the electric field across the ferroelectric is changed with polarization during the transient NC process, rather than a constant. Moreover, based on these experimental results, we calculate the corresponding Gibbs free energy landscape *G* as a function of polarization *P*, as shown in Figures 2 (c) and 2(d). It shows that as *P* increases, *G* first increases and then decreases. This indicates that the system will settle *P* in such a way that does not minimize the *G*. Consequently, the L-K theory, that is, *G* minimization is not applicable in explaining the transient NC. It is necessary to reconsider the thermodynamic process of FE polarization during transient NC.

**3. Thermodynamic driving force of transient NC**

The thermodynamics driving force of transient NC is investigated. Based on the above discussion, we consider the case of $E \neq$ constant. We come back to Equation (5), which is still applicable for the case $E \neq$ constant. According to Equation (5), elastic Gibbs free energy $G_a$ is given by[59, 60]

$$G_a = U - TS - Xx - \frac{1}{2}\varepsilon_0 E^2 \qquad (8)$$
$$= \alpha P^2 + \beta P^4 + \gamma P^6$$

Then, Equation (5) can be rewritten as $dG_a - EdP \leq 0$. Here we define a new thermal potential $\eta$ that satisfies

$$d\eta = dG_a - EdP \leq 0. \qquad (9)$$

This means that the essence of transient NC follows the minimization of the new thermal potential $\eta$. Taking an integration of the above equation yields

$$\int_{\eta_0}^{\eta} d\eta = \int_{G_{a0}}^{G_a} dG' - \int_{P_0}^{P} EdP \leq 0$$

or

$$\eta = G_a - \int_{P_0}^{P} EdP + (\eta_0 - G_{a0}) \qquad (10)$$

where $\eta_0$ and $G_{a0}$ express $\eta$ and $G_a$ at the initial state, respectively. Equation (10) indicates that the new thermal potential $\eta$ is equal to the difference between the elastic Gibbs free energy and electric polarization work. Thus the ferroelectric polarization changes toward the direction where the thermal potential $\eta$ decreases, and this can be equally expressed by

$$\rho \frac{dP}{dt} = -\frac{\partial \eta}{\partial P}. \qquad (11)$$

Taking Equation (10) into Equation (11) gets

$$\rho \frac{dP}{dt} = -\frac{\partial \eta}{\partial P} = E - \frac{dG_a}{dP}. \qquad (12)$$

Combining with Equation (1), we find an interesting result, that is, the expression of $\partial \eta/\partial P$ and $\partial G/\partial P$ is the same. However, thermal potential $\eta$ and $G$ are essentially different because they have completely different physical meanings. Based on the experimental $P$-$E$ curves [see Fig.2(a) and (b)], we calculate the thermal potential $\eta$, as shown in Figures 2(c) and (d). It shows that the $\eta$ exhibits a monotonous decreasing trend. Thus the system will settle $P$ in such a way that minimizes $\eta$ instead of $G$, that is, the thermodynamic driving force of transient NC is the minimization of the $\eta$ instead of $G$.

From Equations (1) and (12), the partial derivative of the Gibbs free energy $G$ with respect to $P$ for a given $V_{FE}$ (or $E$) is equal to the derivative of $\eta$ with respect to the $P$, that is, $d\eta/dP = \partial G/\partial P|_{E=constant}$. This indicates that at a given state (knowing $P$ and $E$ or $V_{FE}$), the polarization switching direction can be judged by the way of minimizing $G$. This means that we could use the Gibbs free energy landscape at a given $E$ to predict the polarization switching direction of the ferroelectric.

## 4. Transient behavior of NC

The dynamic process of polarization switching in the $R$-FEC circuit is further investigated by using numerical simulations. Figure 3 shows the simulated FE voltage response of the $R$-FEC circuit under an applied voltage pulse (-3 V→3 V), based on the Equation (12). The simulation parameters[25] are listed in Table 1. In Figure 3, the states 1 and 18 represent the initial and final stable states, and the states 2~17 express the unstable states. Along the transient polarization states 1→2→3→...→16→17→18, the ferroelectric voltage $V_{FE}$ and ferroelectric polarization $P$ change with time, where the polarization continues to increase, while the ferroelectric voltage increases initially, then decreases, and finally increases again. It can be seen that the NC region appears from state 7 to 12, where $dE/dP<0$.

In order to understand the fundamental physics of the transient NC, we further calculate the Gibbs free energy $G$ landscape as a function of $P$ at different FE voltages, as shown in Figure 4. The $G$ curve is dependent on the electric field across the ferroelectric, and changes as the ferroelectric state changes. Here we define the polarization corresponding to the localized minimum of $G$ as $P_{min}$, which are shown as the black square dots in Figure 4. For the process of states 1→2→3→4→5 ($V_{FE}$: -3→-1.56→-1→0→+1 V), the transient FE polarization state at each ferroelectric voltage is always located on the left side of $P_{min}$. Thus the polarization at these states increases. For the state 6 ($V_{FE}$ = +1.75 V), there are two zero derivative points in the corresponding $G$ curve. One zero derivative point corresponding to negative polarization is not an extreme point (here denoted as $P_{nonmin}$), which means that the energy barrier disappears. At this state, the polarization is localized on the left side of the $P_{nonmin}$, and consequently, the polarization still increases. For the state 7 ($V_{FE}$ = +1.79 V), there is only one zero derivative points, i.e., $P_{min}$, localized at the positive polarization region in the corresponding $G$ curve, as shown in Figure 4. Moreover, there is no energy barrier, and the polarization increases toward $P_{min}$. For the state 8, the $G$ curve is the same as that of case 6. However, the polarization state has been localized on the right side of the $P_{nonmin}$. Thus, for the process of states 6→7→8 ($V_{FE}$: +1.75→+1.79→+1.75 V), the $G$ curve corresponding to each state shows a monotonous decreasing trend and has only one minimum point, which is localized in the positive polarization region (~0.17 C/cm$^2$). These results indicate that the polarization continues to increase. For the process states 9→10→11→12 ($V_{FE}$:

+1→0→-1→-1.56 V), the $G$ curve has the double-well energy landscape, i.e., the energy barrier has appeared. However, the transient polarization state has been always on the right side of the maximum point of the $G$ curve at the corresponding ferroelectric voltage. This means that the transient polarization state has rolled over the energy barrier of the $G$ curve, and the polarization increases toward the right $P_{min}$. This leads to an interesting phenomenon that the polarization for states 7~12 further increases as the voltage decreases, which indicates that the NC phenomenon appears in this process. The process of states 13~18 is similar to states 1~5, that is, the transient polarization state at each ferroelectric voltage is always located on the left side of $P_{min}$, and the polarization increases with the increase of ferroelectric voltage.

The black line in Figure 4 gives the Gibbs free energy during the whole transient NC process. It can be seen that the $G$ is not always decreasing, but changes like an 'M' shape. Above dynamic process reveals that the trend of transient polarization obeys the Gibbs free energy minimization rule under a fixed voltage. However, the whole dynamic process of polarization switching from states 1~18 doesn't satisfy the Gibbs free energy minimization because the voltage is a variable and changed during the transient NC process.

Actually, above dynamic process follows the $\eta$ minimization rule. Figure 5 shows the $G$ and $\eta$ during the transient NC process. The $G$ initially increases, then decreases, then increases, and finally decreases. Thus the $G$ does not always decrease, as expected by the L-K theory. On the other hand, the $\eta$ always decreases. This further confirms our opinion, that is, the thermodynamic driving force of transient NC is indeed the minimization of $\eta$ rather than $G$.

**5. Physical origin of transient NC appearance**

We discuss the fundamental physics of transient NC. Figure 6(a) shows the ideal $P$-$E$ curve obtained by differentiating $G_a$ with respect to $P$. The ideal $P$-$E$ curve is also well known as the 'S' shape curve. Also shown in Figure 6(a) is the real $P$-$E$ curve of transient NC. For each $P$, the $\partial G_a/\partial P$ is equal to $E$ of the ideal $P$-$E$ curve (here denoted as $E_0$). From Equation (12), we can get that the d$P$/d$t$ is proportional to the $E_r$ －$E_0$, where $E_r$ represents the $E$ of transient NC (ideal $P$-$E$ curve). Moreover, the $E_r$－$E_0$ is equal to －$\partial G/\partial P$, which reflect the negative slope of the $G$ vs. $P$ curve. The corresponding $E_r$－$E_0$ profile as a function of $P$ (difference between the black and red lines for each $P$) is given in Fig6(b).

Here we divide the real $P$-$E$ curve into five regions, as shown in Figure 6(a). When $E=E_1$, the energy barrier of the corresponding $G$ curve just disappeared, as shown in Figure 6(c). For region Ⅰ, the $G$ landscape is schematically shown in Figure 6(d). There is only one extreme point, here minimum, and denoted as $P_{min}$. There is no energy barrier, i.e., local maximum. The real $P$ is always localized on the left side of the $P_{min}$. Moreover, the change rate of $P$ (d$P$/d$t$) corresponds to the slope of the $G$ curve, based on Equation (12). And the d$P$/d$t$ is less than the (d$Q$/d$t$), here the $Q$ is the free charges on the metal plate. Thus the d$V_{FE}$/d$t$ is positive, and the $V_{FE}$ increases. For region Ⅱ, the $G$ landscape is schematically shown in Figure 6(e). There are three extreme points, i.e., two local minimum points ($P_{min, L}$ and $P_{min, R}$) and one local maximum point ($P_{max}$). The barrier appears. The real $P$ is always localized on the left side of the $P_{min, L}$. Moreover, the d$P$/d$t$ is less than the (d$Q$/d$t$). Thus the d$V_{FE}$/d$t$ is positive, and the $V_{FE}$ increases. For region Ⅲ (state $a \rightarrow b \rightarrow c$), the $G$ landscape is schematically shown in Figure 6(f). There is only one extreme point ($P_{min}$). The barrier disappears. The real $P$ increases toward the $P_{min}$. From state $a$ to $b$ [see Figure 6(a)], there is no NC phenomenon. From the state $b$ to $c$, the NC phenomenon appears. The d$P$/d$t$ is increasing and becomes larger than d$Q$/d$t$ from state $b$. This is attributed to that the slope of $G$ curve becomes steep from state $b$ to $c$ [see Figure 6(b)]. An interesting phenomenon is that the *dP/dt* before the NC appearance (e.g., *a→b*) is larger than that of state b. This means that the NC is not only dependent on the *dP/dt*. For region Ⅳ (state $c \rightarrow d \rightarrow e$), the $G$ landscape is schematically shown in Figure 6(g). There are three extreme points, i.e., two local minimum points ($P_{min, L}$ and $P_{min, R}$) and one local maximum point ($P_{max}$). The barrier appears. However, the polarization $P$ has rolled over the barrier and quickly increases toward the right $P_{min, R}$. The slope of the $G$ curve in this region is large, especially for the region of state $c$ to $d$. In other words, the distance between the black real *P-E* line and the red ideal *P-E* line is large. This results in larger d$P$/d$t$ than the (d$Q$/d$t$), and consequently NC phenomenon. From the state $d$, the $G$ curve begins to flatten. The d$Q$/d$t$ catches up with the d$P$/d$t$ again, leading to that the NC phenomenon disappears. For region Ⅴ (state $e \rightarrow f \rightarrow g$), the $G$ landscape is schematically shown in Figure 6(h). There is only one extreme point, similar to region Ⅲ [see Figure 6(f)]. The $G$ curve in this region keeps flat and there is still no NC phenomenon. From the above discussion, it can be concluded that the NC appears when the barrier disappears (region $b \rightarrow c$) or the polarization state has rolled over the barrier (region $c \rightarrow d$). There is no case that the polarization climbs up to the barrier top. In addition, the elastic Gibbs free energy $G_a$ profile is shown in Figure 6(i). The polarization states $a \rightarrow b \rightarrow \ldots \rightarrow g$ are also shown. It should be noted that the negative curvature region of $G_a$ is different from the real transient NC region, indicating that the transient NC cannot be judged by the negative curvature of $G_a$.

Based on Equations. (11) and (12), $\partial V_{FE}/\partial t$, at a given time $t_0$, can be expressed as

$$\begin{aligned} \left.\frac{\partial V_{FE}}{\partial t}\right|_{t=t_0} &= \frac{d_{FE}}{\varepsilon_0}\left(\left.\frac{\partial Q}{\partial t}\right|_{t=t_0} - \left.\frac{\partial P}{\partial t}\right|_{t=t_0}\right) \\ &= \frac{d_{FE}}{\varepsilon_0}\left(\frac{V_S - V_{FE_0}}{R} + \left.\frac{\partial G_a}{\rho \partial P}\right|_{t=t_0} - E_0\right) \\ &= \frac{d_{FE}}{\varepsilon_0}\left(\frac{V_S}{R} + \frac{2\alpha P_0 + 4\beta P_0^3 + 6\gamma P_0^5}{\rho} - \left(1 + \frac{d_{FE}}{R}\right)E_0\right) \end{aligned} \quad (13)$$

where $P_0$ and $E_0$ are the polarization and electric field across the ferroelectric at a given time $t_0$, respectively. $d_{FE}$ is the ferroelectric physical thickness. This equation suggests that the appearance of NC is not only related to $V_s$, $R$, $P_0$ and $E_0$, but also related to $\alpha$, $\beta$, $\gamma$ and $\rho$. And these four parameters are related to the ferroelectric material characteristics (intrinsic), while the $V_s$ and $R$ are related to the extrinsic characteristics of the $R$-FEC circuit. This means that the appearance of NC is not only dependent on the ferroelectric material itself but dependent on some extrinsic factors. By further adjusting the value of $\alpha$, $\beta$, and $\gamma$, it is possible to realize that the change of polarization with time is greater than the free charge change with time, leading to the appearance of NC.

In addition, we also explore a physical criterion for transient NC. In the $R$-FEC circuit, electrostatic force work of ferroelectric polarization is given as $W=\int EdP$, and then second-order differentiating $W$ with respect to $P$ is $d^2W/dP^2=dE/dP$. Since $dE/dP<0$ is required for the transient NC, $W$ should satisfy $d^2W/dP^2<0$, which means that the negative curvature of electrostatic force work is a criterion for transient NC. This is obviously different from the traditional criterion that the transient NC region appears in the portion of the negative curvature of elastic Gibbs free energy $G_a$ ($d^2G_a/dP^2<0$). In order to verify these two criteria, we differentiate $G_a$ ($W$) with respect to $P$ and obtain the 'S' shaped (transient) $P$-$E$ curves, as shown in Figure 7. It is found that the transient polarization states in Figure 7 can be directly mapped on the transient $P$-$E$ curve, but not on the 'S' curve. This indicates that the transient NC can indeed be judged by the negative curvature of electrostatic force work instead of elastic Gibbs free energy.

## 4. Conclusion

In conclusion, using the Landau-Ginzburg-Devonshire (L-G-D) theory and the first and second laws of thermodynamics, we have detailedly studied the thermodynamic process of transient NC in the $R$-FEC circuit. Our results show that the dynamic process of polarization switching for transient NC follows the minimization of the new thermodynamic potential $\eta$ (the difference between the elastic Gibbs free energy and the electric polarization work). The conventional view that the Gibbs free energy minimization is not applicable. Furthermore, this viewpoint is verified by the analysis based on experimental *P-E* curves and numerical simulation of the dynamic process of polarization switching for transient NC. In addition, the appearance of the transient NC phenomenon is not due to the widely accepted view that the ferroelectric polarization goes through the negative curvature region of elastic Gibbs free energy landscape ($G_a$). Instead, the transient NC phenomenon appears when the $G_a$ energy tbarrier disappears. Moreover, the transient NC is dependent on both the intrinsic ferroelectric material parameters and extrinsic factors in the *R*-FEC circuit. The work provides a comprehensive explanation for the dynamic process of transient NC.

**Acknowledgments**
This work is supported in part by the National Key Project of Science and Technology of China (Grant no. 2017ZX02315001-002).

Received: ((will be filled in by the editorial staff))
Revised: ((will be filled in by the editorial staff))
Published online: ((will be filled in by the editorial staff))

37. Gao, W. et al. Room-temperature negative capacitance in a ferroelectric–dielectric superlattice heterostructure. *Nano Lett.* **14**, 5814-5819 (2014).

38. Yadav, A. K. et al. Spatially resolved steady-state negative capacitance. *Nature* **565**, 468-471 (2019).

39. Kobayashi, M., Jang, K., Ueyama, N. & Hiramoto, T. Negative capacitance as a performance booster for tunnel FET. In *2016 IEEE Silicon Nanoelectronics Workshop (SNW)* 150-151. (IEEE, 2016).

40. Luk'yanchuk, I., Sene, A. & Vinokur, V. M. Electrodynamics of ferroelectric films with negative capacitance. *Phys. Rev. B* **98**,   (2018).

41. Luk'yanchuk, I., Tikhonov, Y., Sene, A., Razumnaya, A. & Vinokur, V. M. Harnessing ferroelectric domains for negative capacitance. *Communications Physics* **2**, 22 (2019).

42. Hoffmann, M. et al. Direct observation of negative capacitance in polycrystalline ferroelectric $HfO_2$. *Adv. Funct. Mater.* **26**, 8643-8649 (2016).

43. Íñiguez, J., Zubko, P., Luk'yanchuk, I. & Cano, A. Ferroelectric negative capacitance. *Nat. Rev. Mater.* **4**, 243-256 (2019).

44. Liu, Z., Bhuiyan, M. & Ma, T. A critical examination of 'quasi-static negative capacitance'(QSNC) theory. In *2018 IEEE International Electron Devices Meeting (IEDM)* 31.32. 31-31.32. 34. (IEEE, 2018).

45. Van Houdt, J. & Roussel, P. Physical model for the steep subthreshold slope in ferroelectric FETs. *IEEE Electron Device Lett.* **39**, 877-880 (2018).
21

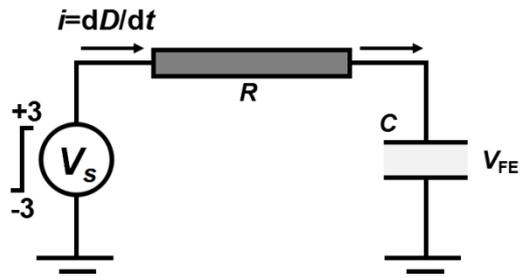

**Figure 1.** The schematic of an *R*-FEC circuit. *C* denotes the FE capacitor and *R* represents the external resistor. *V* and $V_s$ are the voltage across the FE and the applied voltage, respectively.

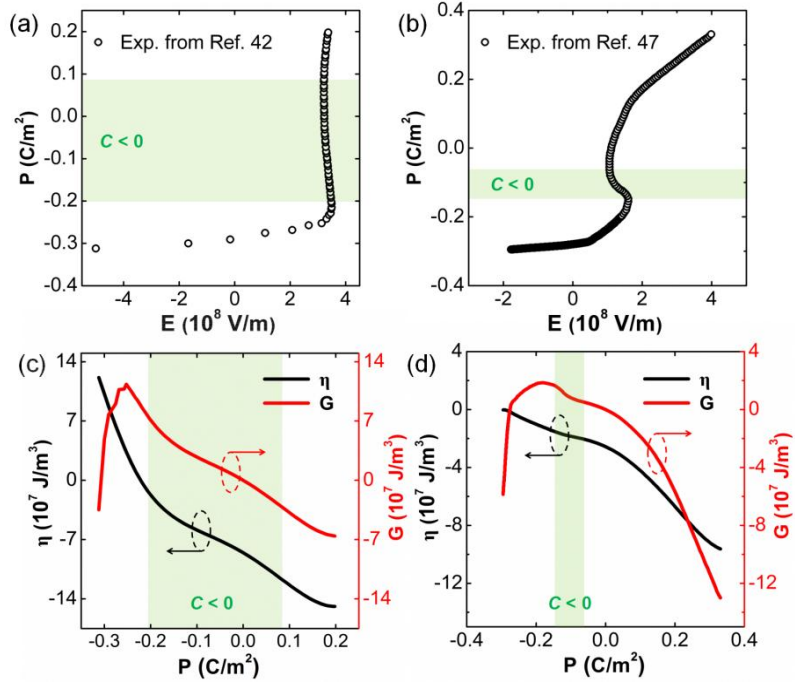

**Figure 2.** (a,b) The experimental *P-E* curves in Refs. 42 and 47. (c,d) The corresponding Gibbs free energy and the new thermal potential $\eta$ profiles as a function of polarization *P*. The shadow regions represent the NC regions.

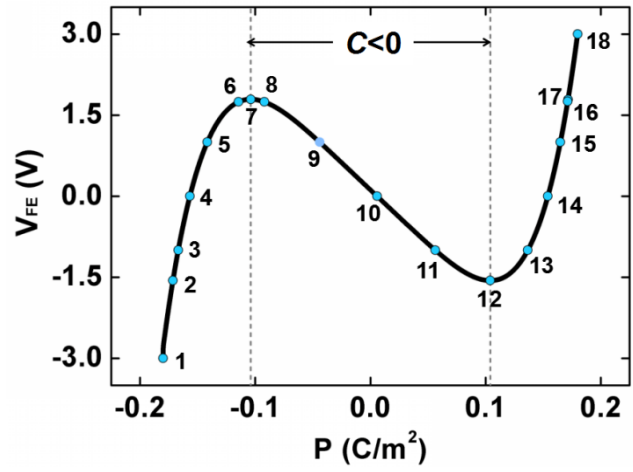

**Figure 3.** The $V_{FE}$-$P$ curve of transient NC. Number 1-18 corresponds to different transient polarization states.

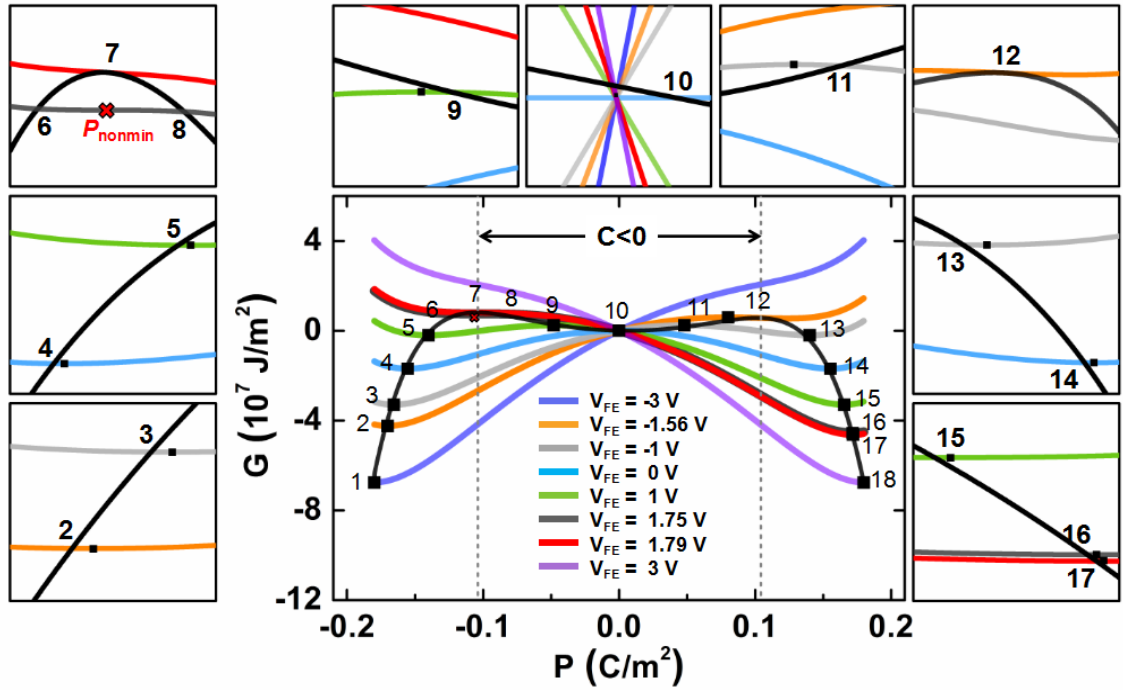

**Figure 4.** The Gibbs free energy profiles as a function of polarization *P* at different voltages for the ferroelectric capacitor. Black square dots represent the local extreme points of the Gibbs free energy at different voltages, respectively. The red cross represents the zero derivative point that is not an extreme point. States 1 and 18 are stable and states 2-17 are unstable states. States 1-18 are connected with the black line. The black line expresses the real *G* in the transient NC process.

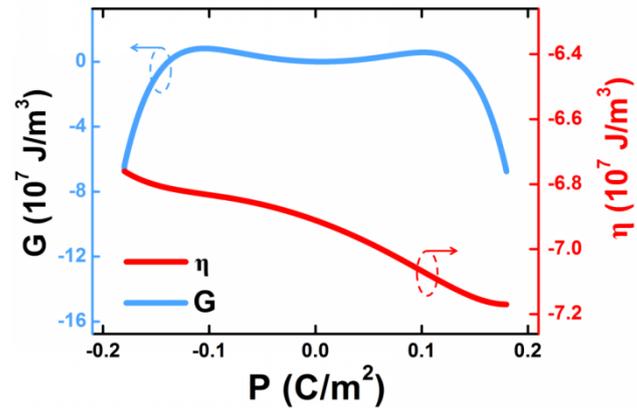

**Figure 5.** The Gibbs free energy profiles and the new thermal potential $\eta$ profiles as a function of polarization $P$.

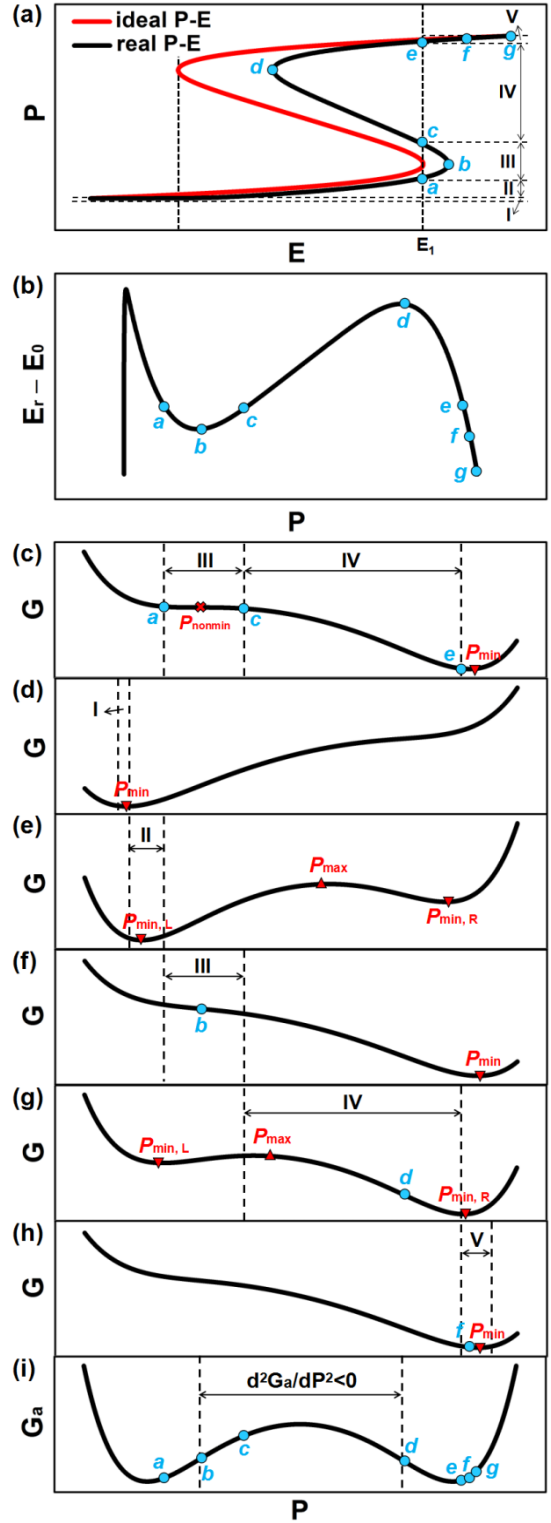

**Figure 6.** (a) The schematic diagram of ideal *P-E* curves and real *P-E* curves. (b) The electric field difference between real *P-E* curves and ideal *P-E* curves as a function of polarization *P*. (c) The Gibbs free energy profile at $E=E_1$. The Gibbs free energy profile for regions I (d), II (e), III (f), IV (g) and V (h). The region III, IV and V correspond to state $a \to b \to c$, $c \to d \to e$ and $e \to f \to g$, respectively. (i) The elastic Gibbs free energy profile as a function of polarization *P*.

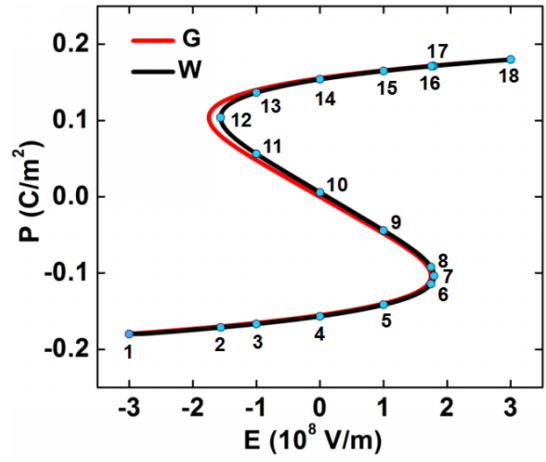

**Figure 7.** The *P-E* curves for elastic Gibbs free energy and electrostatic force work. Number 1-18 corresponds to transient polarization states in Figure 3.

**Table 1.** Simulation parameters for the *R*-FEC circuit.

| Parameter | Quantity | Value[25] |
|---|---|---|
| $\alpha$ | | $-1.05 \times 10^9$ m/F |
| $\beta$ | Landau expansion coefficient | $10^7$ m$^5$/C$^2$F |
| $\gamma$ | | $6 \times 10^{11}$ m$^9$/C$^4$ F |
| $\rho$ | Viscosity coefficient | 500 m sec/F |
| $d_{FE}$ | Ferroelectric thickness | 10 nm |
| $A$ | Capacitor area | $50^2$ μm$^2$ |
| $R$ | External resistance | 50 kΩ |